\documentclass[aps,prfluids,onecolumn,superscriptaddress,longbibliography]{revtex4-2}

\usepackage{color}
\usepackage{graphicx}
\usepackage{siunitx}
\usepackage{amsmath}
\usepackage{hyperref}

\begin{document}

\title{Optimal Turbulent Transport in Microswimmer Suspensions}

\author{Henning Reinken}
\email{henning.reinken@itp.tu-berlin.de}
\affiliation{Institute of Theoretical Physics, Technische Universit\"at Berlin, Stra{\ss}e des 17. Juni 135, 10623, Berlin, Germany} 

\author{Sabine H. L.  Klapp}
\affiliation{Institute of Theoretical Physics, Technische Universit\"at Berlin, Stra{\ss}e des 17. Juni 135, 10623, Berlin, Germany} 

\author{Michael Wilczek}
\affiliation{Theoretical Physics I, University of Bayreuth, Universit\"atsstra{\ss}e 30, 95447, Bayreuth Germany} 

\date{\today}

\begin{abstract}
Microswimmer suspensions self-organize into complex spatio-temporal flow patterns, including vortex lattices and mesoscale turbulence.
Here we explore the consequences for the motion of passive tracers, based on a continuum model for the microswimmer velocity field.
We observe two qualitatively different regimes distinguished via the dimensionless Kubo number $K$.
At advection strengths right above the transition to turbulence, the flow field evolves very slowly ($K \gg 1$) and the spatial vortex structures lead to dominant trapping effects. 
In contrast, deep in the turbulent state, much faster dynamics ($K \ll 1$) consistent with the so-called sweeping hypothesis leads to transport properties completely determined by the temporal correlations.
In between ($K \approx 1$), we observe a regime of optimal transport, signaled by a maximum of the diffusion coefficient.
\end{abstract}


\maketitle

\section{Introduction}
\label{sec: introduction}

One of many interesting phenomena in active fluids~\cite{marchetti2013hydrodynamics,bechinger2016active, doostmohammadi2018active,bar2020self,gompper20202020} is the enhancement of mixing of suspended particles due to the motion of the microwimmers and accompanying hydrodynamic flows.
This effect, observed in recent experiments~\cite{kim2004enhanced,sokolov2009enhanced,leptos2009dynamics,mino2011enhanced,kurtuldu2011enhancement} and simulations~\cite{underhill2008diffusion,ishikawa2010fluid,pushkin2013fluid}, might be utilized, e.g., in microfluidic devices~\cite{kim2007controlled,suh2010review}.
However, it is largely unexplored how the complex emerging large-scale flow patterns, which are typical for such microswimmer suspensions, impact the mixing and transport properties of these non-equilibrium systems.
Here, we quantify active fluid transport in the framework of an experimentally validated model for polar active fluids, which enables a precise control of the flow states ranging from vortex lattices~\cite{wioland2016ferromagnetic,nishiguchi2018engineering,reinken2019anisotropic,reinken2020organizing,reinken2022ising,james2018turbulence,james2021emergence} to active turbulence~\cite{wensink2012meso,bratanov2015new,alert2021active,oza2016generalized,james2018turbulence,james2021emergence}, either externally, e.g.~through obstacles~\cite{nishiguchi2018engineering,reinken2020organizing,reinken2022ising,sone2019anomalous,zhang2020oscillatory}, or by changing the fluid parameters~\cite{reinken2019anisotropic,james2018turbulence,james2021emergence}.

Previous work has mostly focused on characterizing transport deep in the turbulent regime, drawing on analogies either to two-dimensional turbulence \cite{james2018vortex,sanjay2020friction} or to stochastic processes such as L{\'e}vy walks~\cite{mukherjee2021anomalous,ariel2017chaotic}.
Much less is known about transport close to the transition region between different flow states.
Interestingly, equilibrium systems close to structural phase transitions often display anomalous transport properties, e.g., anisotropic and enhanced diffusion near transitions to nematic phases in liquid crystals~\cite{lowen1999anisotropic,lettinga2005self,lettinga2010hydrodynamic} or peaks in the thermal conductivity close to structural changes in crystalline solids~\cite{aramberri2017thermal,chen2018enhancement,chen2019anomalous}.
In contrast, for a nonequilibrium, active system, we rather expect the interplay of the spatial and temporal persistence of flow structures to have a profound impact on transport.
For example, in a stationary vortex lattice, tracer trajectories will be trapped by individual vortices, following closed streamlines. 
This situation is comparable to the transport of electrons in electrostatic potentials, where trajectories are determined by equipotential lines~\cite{reuss1996low,reuss1998percolation,krommes2002fundamental,vlad2004lagrangian,padberg2007lagrangian}.
When the flow becomes turbulent, particles will travel between vortices, giving rise to a diffusive transport for long times, similar to the transport of charged particles in time-varying fields~\cite{isichenko1992percolation,vlad1998diffusion,bakunin2008turbulence}.

Inspired by these analogies, we perform an analysis of the transport properties of active turbulence in terms of the Kubo number $K$, comparing temporal and spatial correlations of the flow field.
We find that the diffusion coefficient as a function of $K$ obtains a maximum close above the transition from a vortex lattice to the turbulent state.
In this regime, the interplay between temporal correlations and spatial structure of the underlying flow leads to optimal transport conditions.

\section{Model}
\label{sec: model}

In this work, we employ an established model for dense microswimmer suspensions~\cite{wensink2012meso,dunkel2013fluid,dunkel2013minimal,bratanov2015new,heidenreich2016hydrodynamic,reinken2018derivation,reinken2019anisotropic,james2018vortex,james2018turbulence} in which density fluctuations can be neglected~\cite{be2020phase} and the dynamics can be described on a coarse-grained (order parameter) level via an effective microswimmer velocity field $\mathbf{v}$~\cite{reinken2018derivation}.
We choose this model over different models that have been shown to exhibit similar pattern formation~\cite{grossmann2014vortex,slomka2017geometry,sone2019anomalous} because it can be derived from microscopic Langevin dynamics including coupling to the solvent flow~\cite{reinken2018derivation} and has been shown to capture experiments on bacteria in unconfined bulk flow~\cite{wensink2012meso} as well as in the presence of geometric confinement, e.g., obstacle lattices~\cite{reinken2020organizing}.
The dynamics of $\mathbf{v}$ is given by
\begin{eqnarray}
\label{eq: dynamic equation}
&\displaystyle\partial_t \mathbf{v} + \lambda \mathbf{v}\cdot\nabla\mathbf{v} = -\frac{\delta \mathcal{F}}{\delta \mathbf{v}},  \\
&\displaystyle \mathcal{F} =  \int d\mathbf{x} \left[q \nabla\cdot\mathbf{v} - \frac{a}{2} |\mathbf{v}|^2 + \frac{b}{4} |\mathbf{v}|^4  +  \frac{1}{2}\left|(1 + \nabla^2)\mathbf{v}\right|^2 \right], \nonumber
\end{eqnarray}
where $q$ is a pressure-like quantity that ensures the incompressibility condition $\nabla \cdot \mathbf{v} = 0$.
The dynamics can be interpreted as a competition between gradient dynamics determined by the functional $\mathcal{F}$ and nonlinear advection ($\lambda \mathbf{v}\cdot\nabla\mathbf{v}$), where $\lambda$ is the advection strength, which can be related to mesoscopic parameters such as the self-propulsion speed~\cite{reinken2018derivation,reinken2022ising}.
For high activity, i.e., $0 < a < 1$, the minimum of $\mathcal{F}$ is a vortex pattern with square lattice symmetry characterized by two perpendicular modes with characteristic wavelength $\Lambda = 2 \pi$ \cite{james2018turbulence}.
When the nonlinear advection strength $\lambda$ is increased above some critical value $\lambda^\star$, which depends on the other parameters $a$ and $b$, the stationary square lattice pattern is destabilized and the advection term induces a dynamic, fluctuating state denoted as mesoscale turbulence~\cite{wensink2012meso,reinken2018derivation}.

We consider $N$ passively advected tracer particles that follow the effective microswimmer velocity field $\mathbf{v}$ governed by Eq.~(\ref{eq: dynamic equation}), according to
\begin{equation}
\label{eq: equation of motion tracers}
\partial_t \mathbf{X}_i (t) = \mathbf{v}(\mathbf{X}_i(t),t) \, ,
\end{equation}
where $\mathbf{X}_i = (X_i,Y_i)$ denotes the position of tracer $i$.
In other words, the tracers sample the Lagrangian trajectories (moving with the flow) of the evolving field $\mathbf{v}(\mathbf{x},t)$.
When considering Eq.~(\ref{eq: equation of motion tracers}) as governing equation for a single tracer, one could argue that it should contain a term stemming from molecular noise, e.g., Brownian diffusion.
However, here we will focus on the impact of advective transport and neglect the influence of molecular diffusion, hence, there is no noise term in Eq.~(\ref{eq: equation of motion tracers}).
To justify this assumption, let us briefly consider the Peclet number $Pe$, which gives the ratio between advective and diffusive transport in a system and is calculated via $Pe = \Lambda_\mathrm{a} v_\mathrm{a} / D_0$, where $\Lambda_\mathrm{a}$ denotes the length scale of the advecting flow, $v_\mathrm{a}$ is the average advective velocity and $D_0$ is the bare molecular diffusion~\cite{bakunin2008turbulence}.
To estimate the Peclet number for tracers in bacterial suspensions exhibiting bacterial turbulence, e.g., \textit{Bacillus Subtilis}~\cite{dunkel2013fluid}, we can first calculate an approximate molecular diffusion coefficient using the Stokes-Einstein relation~\cite{einstein1906theorie}, which is valid for a spherical tracer in three dimensions at low Reynolds number.
The relation is given by $D_0 = k_\mathrm{B}T/(3\pi \eta d)$, where $k_\mathrm{B}$ is the Boltzmann constant, $T$ is the temperature, $\eta$ is the viscosity of the solvent medium and $d$ the diameter of the tracer.
For a tracer of micron size ($d \approx \SI{1}{\micro\meter}$) in water at normal conditions ($T=\SI{293.15}{\K}$, $\eta \approx \SI{0.001}{\newton\second\per\meter\squared}$), we obtain $D_0 \approx \SI{0.4}{\square\micro\meter\per\second}$.
Secondly, the length scale in bacterial turbulence observed in \textit{Bacillus Subtilis} is approximately given by the mean vortex radius $\Lambda_\mathrm{a}\approx\SI{40}{\micro\meter}$~\cite{dunkel2013fluid} and the average velocity $v_\mathrm{a}$ is of the order of $\SI{10}{\micro\meter\per\second}$~\cite{dunkel2013fluid}, depending on the strength of activity.
This estimate yields a Peclet number of $Pe \approx 10^3$, i.e., advective transport is at least three orders of magnitude stronger than the bare molecular diffusion.
These considerations indicate that neglecting molecular diffusion and, thus, a corresponding noise term in Eq.~(\ref{eq: equation of motion tracers}), is indeed a valid assumption, at least for bacterial suspensions exhibiting a mesoscale-turbulent state.
We also neglect any other sources of noise stemming, e.g., from chemical heterogeneities.

We employ a pseudo-spectral method to solve Eq.~(\ref{eq: dynamic equation}) in a two-dimensional system with periodic boundary conditions starting from random initial values and simultaneously evolve the tracer trajectories according to Eq.~(\ref{eq: equation of motion tracers}), see Appendix~\ref{app: numerical methods} for details on the numerical methods.

\section{Transition to turbulence}
\label{sec: transition}

\begin{figure}
\includegraphics[width=0.99\linewidth]{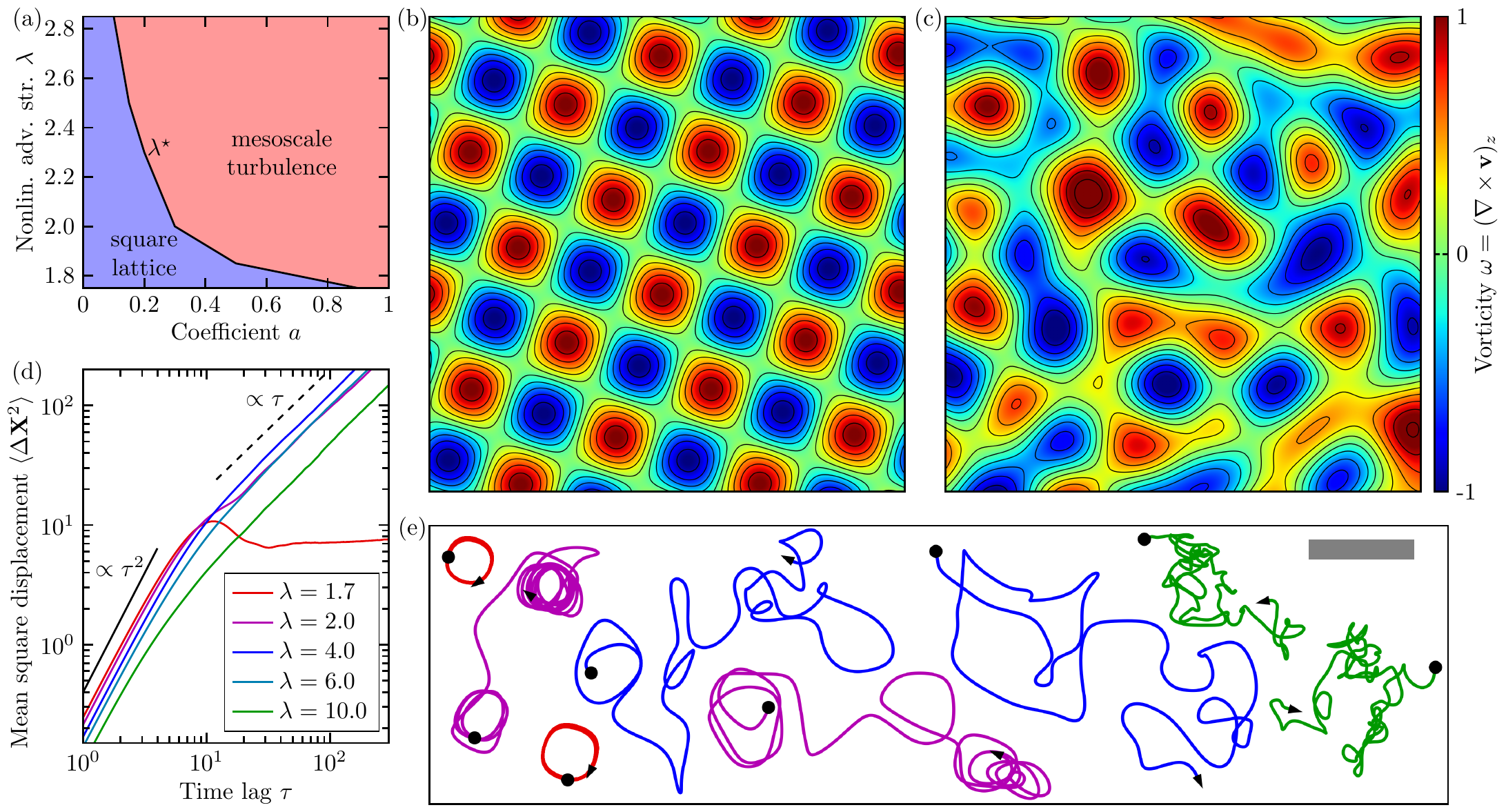}
\caption{\label{fig: transition}Transition to turbulence. (a) Numerically obtained threshold value of the nonlinear advection strength $\lambda^\star$ as a function of $a$ ($b=1.6$). Below $\lambda^\star$, the system settles into a stationary vortex lattice, whereas above $\lambda^\star$, a mesoscale-turbulent state emerges. Snapshots of the vorticity field $\omega = (\nabla \times \mathbf{v})_z$ at $a=0.5$ in the lattice state at (b) $\lambda=1.7$ and in the mesoscale-turbulent state at (c) $\lambda=4$. Tracers move along instantaneous streamlines (solid lines) of the flow, which continuously break up and reconnect in the turbulent state. (d) MSD as a function of time lag $\tau$ for different $\lambda$ and $a=0.5$ (e) Sample trajectories for $\lambda=1.7$ (red), $\lambda = 2$ (violet), $\lambda = 4$ (blue) and $\lambda = 10$ (green) and elapsed time $\Delta t = 250$. The same scale is used for (b), (c) and (e), where the gray bar indicates a length of $2\pi$.}
\end{figure}

We first discuss spatial and temporal correlations of $\mathbf{v}(\mathbf{x},t)$, which are characterized in the Eulerian frame (fixed in space). 
Note that we make the assumption of statistically homogeneous, stationary and isotropic turbulence throughout this work.
Thus, the correlation functions do not depend on space $\mathbf{x}$, time $t$ and orientation, but only on the distance $r$ and time lag $\tau$.
The longitudinal correlation function $f(r)$~\cite{davidson2015turbulence} is defined via
\begin{equation}
\label{eq: longitudinal correlation function}
f(r) = v^{-2}\langle v_x(\mathbf{x},t) v_x(\mathbf{x}+r\mathbf{e}_x, t) \rangle \, ,
\end{equation}
where $\mathbf{e}_x$ denotes the unit vector in $x$-direction and the average $\langle \dots \rangle$ is performed over all $\mathbf{x}$ and $t$.
Further, the component-wise root-mean-square velocity $v$, defined via $v^2 = \langle v_x^2 \rangle = \langle v_y^2\rangle$, gives a measure of the overall strength of the flow field.
The Eulerian temporal correlation function $C_\mathrm{E}(\tau)$ is defined via 
\begin{equation}
\label{eq: Eulerian temporal correlation function}
C_\mathrm{E}(\tau) = \langle \mathbf{v}(\mathbf{x},t)\cdot \mathbf{v}(\mathbf{x}, t+\tau) \rangle \, .
\end{equation}
Integrating the correlation functions over $r$ and $\tau$, respectively, we define a characteristic length and time scale of the evolving field $\mathbf{v}(\mathbf{x},t)$~\cite{davidson2015turbulence}.
In particular, the longitudinal correlation length $\ell$ and the Eulerian correlation time $\tau_\mathrm{E}$ are calculated via
\begin{equation}
\label{eq: integral length scale and Eulerian correlation time}
\ell = \int_0^\infty f(r)\, dr \quad \text{and} \quad 
\tau_\mathrm{E} = \frac{1}{2v^2}\int_0^\infty C_\mathrm{E}(\tau) \, d\tau \, .
\end{equation}
The field $\mathbf{v}(\mathbf{x},t)$ displays two types of spatio-temporal structures depending on the nonlinear advection strength $\lambda$.
In particular, there is a threshold value $\lambda^\star$, below which the system settles into a regular square vortex lattice (see also Appendix~\ref{app: vortex lattice}) given as the minimum of the functional $\mathcal{F}$ in Eq.~(\ref{eq: dynamic equation}).
In contrast, for $\lambda > \lambda^\star$, this stationary, non-fluctuating state is destabilized and the system exhibits a dynamic, mesoscale-turbulent state, see Fig.~\ref{fig: transition} for snapshots.
When approaching the threshold value $\lambda^\star$ from above, the Eulerian correlation time will diverge due to the development of a non-zero tail of $C_\mathrm{E}(\tau)$ (see Appendix~\ref{app: Eulerian correlations}).
The value $\lambda^\star$ depends on the other parameters and can be obtained numerically by determining when a stationary square lattice emerges.
Fig.~\ref{fig: transition}(a) shows the value of $\lambda^\star$ as a function of the coefficient $a$ of the linear term in Eq.~(\ref{eq: dynamic equation}) at $b=1.6$.
For the values of $a$ considered in this work, the threshold value is in the range $\lambda^\star \approx 1.7 \dots 2.8$.

\section{Transport properties}
\label{sec: transport properties}

The emerging states of the flow field $\mathbf{v}(\mathbf{x},t)$ determine the shape of the tracer trajectories $\mathbf{X}_i(t)$ via Eq.~(\ref{eq: equation of motion tracers}).
In particular, in the stationary vortex lattice below $\lambda^\star$, the tracers move along closed loops, whereas above $\lambda^\star$, the trajectories become increasingly irregular.
The resulting transport behavior is quantified by the mean squared displacement (MSD),
\begin{equation}
\label{eq: MSD}
\langle \Delta \mathbf{X}^2 \rangle(\tau) = \frac{1}{N} \sum_{i=1}^N \big(\mathbf{X}_i(t_0 +\tau) - \mathbf{X}_{i}(t_0)\big)^2 \, ,
\end{equation}
where $\mathbf{X}_{i}(t_0)$ is the initial position of tracer $i$ at time $t_0$.
In Fig.~\ref{fig: transition}(d), the MSD is plotted as a function of time lag $\tau$ at different values of $\lambda$, while
corresponding sample trajectories are shown in (e).
Initially, we observe ballistic behavior, i.e., $\langle \Delta \mathbf{X}^2 \rangle \propto \tau^2$.
For longer time lags, the behavior depends on the type of emerging state.
Below $\lambda^\star$, due to the trapping of tracers in closed loops within vortices of the stationary lattice, see Fig.~\ref{fig: transition}, the MSD will saturate at a constant value.
In contrast, for $\lambda > \lambda^\star$, the emerging turbulent state allows the tracers to escape closed loops, introducing randomness to the trajectories, see Fig.~\ref{fig: transition}(e).
As a consequence, the behavior becomes diffusive after the initial ballistic time scale.
The slope $\langle \Delta \mathbf{X}^2 \rangle(\tau)$ is proportional to the diffusion coefficient $D$, defined analogously to Brownian motion, i.e., $\langle \Delta \mathbf{X}^2 \rangle(\tau) = 2 d D \tau$, where $d$ denotes the spatial dimension ($d=2$ in our case).
Upon further increase of $\lambda$, we observe another effect:
As tracers are transported with $\mathbf{v}$, but the structures in the flow field itself with $\lambda \mathbf{v}$ [see Eq.~(\ref{eq: dynamic equation})], the tracer motion increasingly decouples from that of the flow field, see Supplementary Movies.
As a result, the ballistic time scale decreases, trajectories become more irregular, see Fig.~\ref{fig: transition}(e), and the long-time MSD shifts towards smaller values.

\begin{figure}
\includegraphics[width=0.99\linewidth]{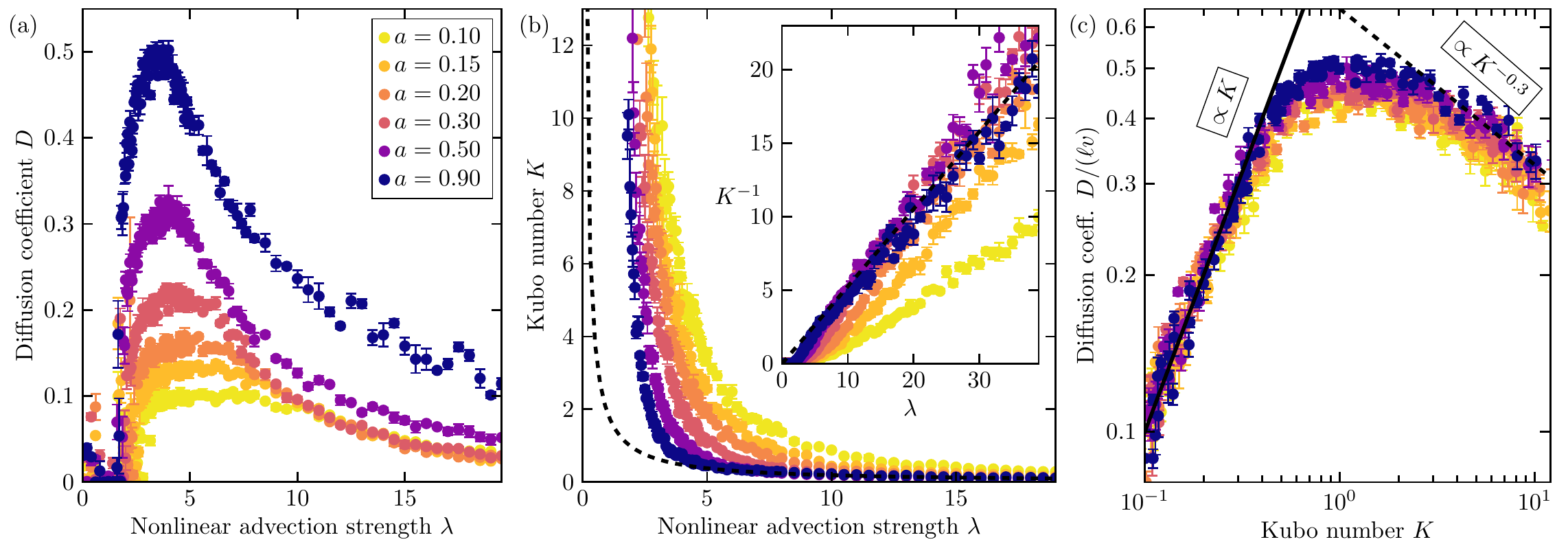}
\caption{\label{fig: diffusion}(a) Diffusion coefficient $D$ as a function of $\lambda$ for different values of $a$ ($b=1.6$). (b) Kubo number $K$ as a function of $\lambda$. Inset: inverse Kubo number $K^{-1}$ as a function of $\lambda$. The dashed line gives the result obtained via Kraichnan's random sweeping hypothesis with $c = 0.33$, see Eq.~(\ref{eq: inverse Kubo number sweeping}). (c) Dimensionless diffusion coefficient $D/(\ell v)$ as a function of $K$. The scaling in the different regimes is given as solid and dashed line, respectively. Error bars represent the standard error.}
\end{figure}

Aiming for a more systematic analysis, we plot the diffusion coefficient in Fig.~\ref{fig: diffusion}(a) as a function of $\lambda$ for different values of $0 < a < 1$.
After an initial trapping regime, where $D = 0$, diffusion increases rapidly above $\lambda^\star$ for all values of $a$.
Strikingly, we observe a clear maximum at $\lambda > \lambda^\star$ in an intermediate regime, $3 < \lambda < 6$, above which $D$ slowly decreases.
Further note that an increase of the activity (measured by the coefficient $a$) consistently leads to higher $D$ due to enhancement of the mean kinetic energy~$\propto v^2$ of the flow, compare Eq.~(\ref{eq: diff coeff Lagrange}) in Appendix~\ref{app: Lagrangian correlations}.

To unravel the nonmonotonic behavior of $D(\lambda)$ and, in particular, the location of the maximum, we take a closer look at the subtle interplay between different time scales of the flow field.
To this end, we will borrow a concept commonly used for transport problems in random fields, e.g., electrons in turbulent magnetized plasmas~\cite{reuss1996low,reuss1998percolation,vlad1998diffusion,padberg2007lagrangian}: the dimensionless Kubo number $K$~\cite{kubo1963stochastic,krommes2002fundamental,bakunin2008turbulence} defined via
\begin{equation}
\label{eq: Kubo number}
K = \frac{\tau_\mathrm{E}}{\tau_\mathrm{tr}} \, .
\end{equation}
It compares the Eulerian correlation time $\tau_\mathrm{E}$ [Eq.~(\ref{eq: integral length scale and Eulerian correlation time})] with the transport time scale $\tau_\mathrm{tr} = \ell/v$, often denoted as large eddy turnover time in inertial turbulence~\cite{davidson2015turbulence}.
We recall that $\ell$ is completely determined by spatial correlations of the advecting flow. 
Thus, $\tau_\mathrm{tr}$ measures how long it takes a tracer to be transported a distance $\ell$ by $\mathbf{v}(\mathbf{x},t)$.
For fields evolving very quickly compared to $\tau_\mathrm{tr}$, we have $K \ll 1$, whereas for slowly evolving fields, we have $K \gg 1$.
Fig.~\ref{fig: diffusion}(b) shows $K$ as a function of $\lambda$.
The divergent behavior when approaching the transition to the stationary state at $\lambda^\star$ is inherited from the likewise diverging Eulerian correlation time.
Remarkably, we observe a proportionality $K \propto \lambda^{-1}$ for larger advection strength, i.e., $K^{-1} \propto \lambda$, see inset.
The plausibility of this linear relation can be shown on theoretical grounds using Kraichnan's idealized random sweeping hypthesis~\cite{kraichnan1964kolmogorov,tennekes1975eulerian,wilczek2012wave} based on the assumption that decorrelation processes are dominated by the sweeping of small-scale structures by the large-scale flow.
In mathematical terms, this can be written as an advection problem,
\begin{equation}
\label{eq: dynamic equation linearized sweeping}
\partial_t \mathbf{v} \approx - \lambda \mathbf{v}_\mathrm{s} \cdot \nabla  \mathbf{v}\, ,
\end{equation}
where we have modified the original ansatz~\cite{kraichnan1964kolmogorov,wilczek2012wave} by adding the advection strength $\lambda$ as a prefactor on the right-hand side, motivated by the structure of Eq.~(\ref{eq: dynamic equation}).
As the fluctuations of the sweeping velocity field $\mathbf{v}_\mathrm{s}$ ultimately stem from the flow field $\mathbf{v}$, it is reasonable to assume that the sweeping velocity variance $v_\mathrm{s}^2$ is proportional to the variance of $\mathbf{v}$ given by $v^2$.
Here, we introduce the proportionality constant $c$ via $v_\mathrm{s} = c\, v$, to be determined by comparison with numerical results.
Since the calculation is quite involved, we detail all the steps in appendices~\ref{app: sweeping} and \ref{app: connection energy spectrum integral length} and only state the final result here:
\begin{equation}
\label{eq: inverse Kubo number sweeping}
K^{-1} = \frac{4c\lambda}{\sqrt{2\pi}} \propto \lambda \, .
\end{equation}
As seen from the inset of Fig.~\ref{fig: diffusion}(b), the sweeping hypothesis is remarkably accurate for the highly dynamic, turbulent flow field at larger values of both $\lambda$ and $a$.
Via comparison with numerical results, we find $c \approx 0.33$ for the proportionality constant.
Surprisingly, this value seems to be quite universal across wide ranges of parameter sets and does not depend on $a$ and $b$ as long as $a > 0.2$ (see Appendix~\ref{app: variation of b} for a variation of $b$).
The resulting linear relation is shown in Fig.~\ref{fig: diffusion} as the dashed line.
As only part of the energy of the flow field resides in the large-scale structures responsible for the sweeping effect, a value $c < 1$ seems indeed plausible from an intuitive point of view.
Closer to the transition, where the dynamics becomes much slower, the sweeping effect does not seem to be the dominant driver of temporal decorrelation and more subtle interactions between pattern formation and nonlinear advection evidently play a larger role.

Having introduced the Kubo number, we are now equipped to characterize different regimes of transport.
For smaller values of $\lambda$ close to the transition at $\lambda^\star$, the flow field evolves very slowly, i.e., $\tau_\mathrm{E} \gg \tau_\mathrm{tr}$, which yields $K \gg 1$, see Fig.~\ref{fig: diffusion}(b). 
In this regime, vortices persist for long times before moving or vanishing, which means that tracers are frequently trapped and their trajectories describe full circular orbits for multiple rotations before being transported further away, see also Fig.~\ref{fig: transition}(e).
Increasing $\lambda$ leads to faster dynamics of the flow field, the trapping effect becomes less dominant, and the diffusion coefficient increases as is shown in Fig.~\ref{fig: diffusion}(a).
When the maximum of $D$ is reached, we have $K \approx 1$. 
Here, the two time scales $\tau_\mathrm{E}$ and $\tau_\mathrm{tr}$ become comparable, which means that it takes approximately the same amount of time for the flow field to rearrange itself as it takes a tracer to move to the other side of a vortex comparable in size to $\ell$.
Before the orbit can transport the tracer back to its original position, $\mathbf{v}(\mathbf{x},t)$ has changed considerably and the original vortex has moved or vanished.
This interplay between timescales creates optimal transport conditions.
Increasing $\lambda$ even further, we observe a third regime where $D$ decreases.
Here, $\tau_\mathrm{E} \ll \tau_\mathrm{tr}$ (yielding $K \ll 1$) and tracers are not able to reach the other side of a vortex before rearrangement.
The Lagrangian correlation time $\tau_\mathrm{L}$ approaches the Eulerian correlation time $\tau_\mathrm{E}$ and the spatial structure of $\mathbf{v}(\mathbf{x},t)$ becomes increasingly unimportant for turbulent diffusion.
As a consequence, the diffusion coefficient scales as $D\propto\tau_\mathrm{E} \propto K$, see Appendix~\ref{app: Lagrangian correlations} for details.
Indeed, for particle transport by a random potential~\cite{isichenko1992percolation,vlad2004lagrangian,padberg2007lagrangian}, this proportionality is an established result and usually denoted as quasi-linear regime, valid for $K \ll 1$.

To investigate the scaling behavior further, we plot the nondimensionalized diffusion coefficient $D/(\ell v)$ in a log-log plot as a function of $K$, see Fig.~\ref{fig: diffusion}(c).
After the quasi-linear regime for $K \ll 1$ (larger $\lambda$), where $D \propto K$, we find the maximum of $D$ at $K \approx 1$.
Then, the diffusion coefficient decreases for $K \gg 1$ (smaller $\lambda$) due to the dominance of trapping effects. 
To understand this high-Kubo-number regime, we have to remind ourselves that transport of tracers always follows the instantaneous streamlines of the flow field, see Eq.~(\ref{eq: equation of motion tracers}).
As these persist for long times due to the slow dynamics, the main ``drivers'' of transport are those streamlines in between vortices that continue for large distances and do not curve back on themselves, compare Fig.~\ref{fig: transition}(c).
This picture is reminiscent of percolation theory, on the basis of which a scaling law can be derived analytically for transport in random potentials acting like stream functions~\cite{gruzinov1990two}.
The derived scaling is $D \propto K^{-0.3}$, which has also been confirmed numerically~\cite{reuss1996low,reuss1998percolation,vlad1998diffusion}.
Remarkably, the scaling is also consistent with our results as shown in Fig.~\ref{fig: diffusion}(c).
We stress that this scaling is robust against variation of the coefficient $b$, see Appendix~\ref{app: variation of b}.

\section{Conclusions}
\label{sec: conclusions}

We have investigated transport of particles in active turbulence using a continuum model for polar active fluids, which exhibits a range of flow states.
Maximal diffusion coefficients are reached for intermediate advection parameters slightly above the transition from a regular vortex lattice to active turbulence.
Drawing on analogies to transport in random fields and, in particular, magnetized plasmas, we borrow the concept of the Kubo number and show that this optimal turbulent transport occurs at $K \approx 1$, where the flow balances spatio-temporal persistence and dynamics.
Additionally, we rationalize the Kubo number scaling for large active advection using Kraichnan's random sweeping hypothesis~\cite{kraichnan1964kolmogorov,wilczek2012wave}, establishing analogies to transport in classical hydrodynamic turbulence.

From a more general perspective, our work describes a novel, striking example of how a non-equilibrium transition between different collective states of an active system leads to a diffusion anomaly. 
While such anomalies are well established in the context of structural phase transitions of {\em equilibrium} systems, e.g., liquid crystals \cite{lowen1999anisotropic,lettinga2005self,lettinga2010hydrodynamic}, corresponding non-equilibrium effects are less explored and often restricted to simple models such as driven Brownian particles~\cite{reimann2001giant,goychuk2021nonequilibrium} or lattice models~\cite{hinrichsen2000non}. 
In contrast, here we have established a clear link between a large-scale pattern formation phenomenon in a generic continuum model of a polar active fluid, and a diffusion maximum. 

As discussed in section~\ref{sec: model}, we neglect the influence of molecular diffusion as we expect the diffusion coefficient $D_0$ to be much smaller than the turbulent diffusion coefficient observed in this study.
We note that for transport in two-dimensional \textit{stationary fields}, molecular diffusivity independent of advection is indeed essential to facilitate transport by letting tracers escape from otherwise closed streamlines~\cite{bakunin2008turbulence}.
Such a \textit{seed} diffusivity $D_0$ would of course alter the behavior in the stationary vortex lattice state and for very high Kubo numbers, where trapping effects play a significant role.
However, in the region of optimal turbulent transport, where the maximum of the diffusion coefficient is reached, the flow field already rearranges itself quite quickly and trapping effects do not play a dominant role.
Thus, a strong impact of molecular diffusion on the behavior in this region is not expected.
Still, a systematic investigation of the influence of (molecular) noise on the tracer dynamics is an interesting subject for future studies.

Further, starting from the present findings for passive point-like tracers, it seems promising to investigate additional effects such as impact of inertia, shape or activity of the tracers~\cite{li2020diffusion}. 
Indeed, for passive flows, such aspects have already been explored in some detail, e.g.,~\cite{pandit2017overview,voth2017anisotropic}. 
Active flows provide an intriguing generalization of such studies, whose implications for phenomena like mixing have yet to be explored.

\begin{acknowledgments}
This work was funded by the Deutsche Forschungsgemeinschaft (DFG, German Research Foundation) - Projektnummer 163436311 - SFB 910.
\end{acknowledgments}

\appendix

\section{Numerical methods}
\label{app: numerical methods}

For the numerical simulations we use the vorticity formulation, which is obtained by taking the curl of Eq.~(1), yielding
\begin{equation}
\label{eq: vorticity formulation}
\partial_t \omega + \lambda \mathbf{v}\cdot\nabla\omega = a \omega  - b \big[\nabla \times (|\mathbf{v}|^2\mathbf{v})\big]_z   - (1 + \nabla^2)^2 \omega \, ,
\end{equation}
where $\omega = (\nabla \times \mathbf{v})_z$ denotes the vorticity in the vertical direction.
The spatially constant mean velocity $\langle \mathbf{v} \rangle$, which is not contained in the vorticity field, is evolved simultaneously via  
\begin{equation}
\label{eq: dynamic equation for homogeneous contribution}
\partial_t \langle\mathbf{v}\rangle= (a - 1) \langle \mathbf{v} \rangle - b \langle |\mathbf{v}|^2 \mathbf{v} \rangle \, .
\end{equation}
Here, $\langle \dots \rangle$ denotes the spatial average.
The full spatially dependent velocity field is then obtained using the stream function $\Phi$ via
\begin{equation}
\label{eq: velocity from stream function}
\mathbf{v} = \langle \mathbf{v} \rangle + \begin{pmatrix} - \partial_y \Phi \\ \hphantom{-} \partial_x \Phi \end{pmatrix} \, ,
\end{equation}
which automatically fulfills the incompressibility conditon $\nabla \cdot \mathbf{v} = 0$.
The stream function and the vorticity are connected via
\begin{equation}
\label{eq: stream function from vorticity}
\nabla^2 \Phi = \omega \, .
\end{equation}
The dynamics is calculated in a two-dimensional box of side length $L$, where the spatial dimensions are discretized by $n\times n$ grid points.
For reasons of accuracy, we employ a pseudo-spectral method. 
For the time integration we use Euler's method combined with an operator splitting technique to solve the linear part of Eq.~(\ref{eq: vorticity formulation}) exactly.

Note that the stream function $\Phi(x,y,t)$ is analogous to the fluctuating potential introduced to investigate transport properties in random fields, e.g., electrons in plasma physics~\cite{reuss1996low,reuss1998percolation,vlad1998diffusion,krommes2002fundamental,padberg2007lagrangian}.
The tracers move along equipotential lines of $\Phi(x,y,t)$ and their equations of motion can also be written as
\begin{equation}
\label{eq: tracer dynamics from stream function}
\begin{aligned}
\partial_t X_i (t) &= - \partial_y \Phi(X_i,Y_i,t)\, ,\\
\partial_t Y_i (t) &= \hphantom{-} \partial_x \Phi(X_i,Y_i,t) \, .
\end{aligned}
\end{equation}

In order to determine the motion of tracer particles according to Eq.~(\ref{eq: equation of motion tracers}) or, equivalently Eq.~(\ref{eq: tracer dynamics from stream function}), we need the Eulerian velocity field $\mathbf{v}(\mathbf{x},t)$ at arbitrary points in continuous space $\mathbf{x}$.
To interpolate the field $\mathbf{v}(\mathbf{x},t)$ from the velocity field in the discretized box coordinates (discontinuous space) we employ bicubic splines.
The trajectories of the tracers are calculated via a four-step Adams–Bashforth method, which allows for higher accuracy than Euler's method without losing computational efficiency.
Before starting the analysis of the tracer trajectories, we let the dynamics of $\mathbf{v}$ evolve for at least $500$ time units to achieve a statistically steady state.
The tracers are then added to the calculation using initial spatial coordinates taken from a uniform distribution covering the whole box.

\section{Stationary vortex lattice}
\label{app: vortex lattice}

For small advection strength, $\lambda < \lambda^\star$, the system settles into a stationary square vortex lattice state characterized by the wavenumber $k_\mathrm{c} = 1$.
We make the following ansatz for such a vortex lattice,
\begin{equation}
\begin{aligned}
v_x(x,y) = \hat{v} \cos(k_\mathrm{c} y)\, ,\\
v_y(x,y) = \hat{v} \cos(k_\mathrm{c} x)\, ,
\end{aligned}
\label{eq: vortex lattice}
\end{equation}
where $\hat{v}$ is a velocity amplitude.
Note that Eq.~(\ref{eq: vortex lattice}) fulfills the  incompressibility condition $\nabla \cdot \mathbf{v} = 0$.
For a velocity field according to Eq.~(\ref{eq: vortex lattice}), the tracer trajectories $\mathbf{X}(t)$ are given as solutions to the following initial value problem,
\begin{equation}
\begin{aligned}
\partial_t X(t) &= \hat{v}\cos(k_\mathrm{c}Y(t))\, ,\\
\partial_t Y(t) &= \hat{v}\cos(k_\mathrm{c}X(t))\, ,\\
X(0) &= X_0\, ,\\
Y(0) &= Y_0\, .
\end{aligned}
\label{eq: initial value problem vortex lattice}
\end{equation}
Using the method of separation of variables, we obtain a condition for the closed trajectories of the tracers depending on the initial position $\mathbf{X}_0 = (X_0,Y_0)$, 
\begin{equation}
\label{eq: trajectories of tracers vortex lattice}
\sin(k_\mathrm{c}X) - \sin(k_\mathrm{c}X_0) = \sin(k_\mathrm{c}Y) - \sin(k_\mathrm{c}Y_0)\, .
\end{equation}
Below the critical value of the nonlinear advection strength, $\lambda < \lambda^\star$, tracers must follow trajectories according to Eq.~(\ref{eq: trajectories of tracers vortex lattice}).

Further, we can calculate the Eulerian spatial correlation functions $f(r)$ and $C_\mathrm{E}(r)$ [see Eqs.~(\ref{eq: longitudinal correlation function}) and (\ref{eq: spatial Eulerian velocity correlation f})] for a velocity field given by Eq.~(\ref{eq: vortex lattice}).
The calculation is straightforward and yields
\begin{equation}
\begin{aligned}
f(r) &= J_0(k_\mathrm{c}r) + J_2(k_\mathrm{c}r)\, ,\\
C_\mathrm{E}(r) &= 2 v^2 J_0(k_\mathrm{c}r)\, ,
\end{aligned}
\label{eq: Eulerian spatial correlations vortex lattice}
\end{equation}
where $J_n(r)$ denotes the $n$-th Bessel function of first kind.
Using the result for $f(r)$, the correlation length can be calculated exactly for a regular vortex lattice via Eq.~(\ref{eq: integral length scale and Eulerian correlation time}), which yields 
\begin{equation}
\label{eq: correlation length vortex lattice}
\ell = \int_0^\infty J_0(k_\mathrm{c}r) + J_2(k_\mathrm{c}r) \, dr = 2/k_\mathrm{c}\, .
\end{equation}
Setting $k_\mathrm{c} = 1$, we have $\ell = 2$, which is also observed in Fig.~\ref{fig: Eulerian quan}(b) for $\lambda < \lambda^\star$.

\section{Eulerian correlations}
\label{app: Eulerian correlations}

In this section, we discuss spatial and temporal correlations in the Eulerian framework.
To this end, let us first introduce a few important quantities and definitions (for details, see Refs.~\cite{wilczek2012wave,davidson2015turbulence}).
We start with the Fourier transform (with respect to space) of the velocity field, which can be calculated in 2D via
\begin{equation}
\label{eq: Fourier transform velocity field}
\begin{aligned}
\mathbf{v}(\mathbf{k},t) &= \frac{1}{4 \pi^2}\int d\mathbf{x}\, \mathbf{v}(\mathbf{x},t)\exp(-i\mathbf{k}\cdot\mathbf{x})\, ,\\
\mathbf{v}(\mathbf{x},t) &= \int d\mathbf{k}\, \mathbf{v}(\mathbf{k},t)\exp(i\mathbf{k}\cdot\mathbf{x})\, .
\end{aligned}
\end{equation}
The full Eulerian velocity correlation tensor is calculated via
\begin{equation}
\label{eq: velocity covariance tensor}
C^\mathrm{E}_{ij}(\mathbf{r},\tau) = \big\langle v_i(\mathbf{x},t) v_j (\mathbf{x} + \mathbf{r},t + \tau)\big\rangle\, ,
\end{equation}
where, in statistically stationary and homogeneous turbulence, $C^\mathrm{E}_{ij}$ only depends on the distance vector $\mathbf{r}$ and time lag $\tau$ and is independent of the coordinates $\mathbf{x}$ and $t$.
The Fourier transform of $C^\mathrm{E}_{ij}(\mathbf{r},\tau)$ is denoted as time-lag-dependent energy spectrum tensor $E_{ij}(\mathbf{k},\tau)$, 
\begin{equation}
\label{eq: two-time energy spectrum tensor}
E_{ij}(\mathbf{k},\tau) = \frac{1}{4 \pi^2}\int d\mathbf{r}\, C^\mathrm{E}_{ij}(\mathbf{r},\tau)\exp(-i\mathbf{k}\cdot\mathbf{r})\, ,
\end{equation}
\begin{equation}
\label{eq: two-time energy spectrum tensor transform}
C^\mathrm{E}_{ij}(\mathbf{r},\tau) = \int d\mathbf{k}\, E_{ij}(\mathbf{k},\tau)\exp(i\mathbf{k}\cdot\mathbf{r})\, .
\end{equation}
The instantaneous energy spectrum tensor $E_{ij}(\mathbf{k})$ is defined analogously as the Fourier transform of the instantaneous correlation tensor $C^\mathrm{E}_{ij}(\mathbf{r},\tau = 0) = C^\mathrm{E}_{ij}(\mathbf{r})$.
In two-dimensional isotropic turbulence, the full energy spectrum tensor can be written as
\begin{equation}
\label{eq: energy spectrum tensor wavenumber}
E_{ij}(\mathbf{k}) = \frac{\tilde{E}(k)}{\pi k} \left( \delta_{ij} - \frac{k_i k_j}{k^2}\right)\, ,
\end{equation}
where we have introduced the scalar energy spectrum $\tilde{E}(k)$, which is just a function of the wavenumber $k = |\mathbf{k}|$.
The energy spectrum is related to the trace of the full tensor $E_{ij}(\mathbf{k})$ via
\begin{equation}
\label{eq: energy spectrum wavenumber}
\tilde{E}(k) = \pi k E_{ii}(\mathbf{k}) \, .
\end{equation}
As a result of this definition, the mean kinetic energy is given as the integral
\begin{equation}
\label{eq: integral energy spectrum wavenumber}
\frac{1}{2} \langle |\mathbf{v}|^2 \rangle = v^2  = \int dk\, \tilde{E}(k)  \, .
\end{equation}
The time-lag-dependent quantity $\tilde{E}(k,\tau)$ is defined analogously.

\begin{figure}
\includegraphics[width=0.99\linewidth]{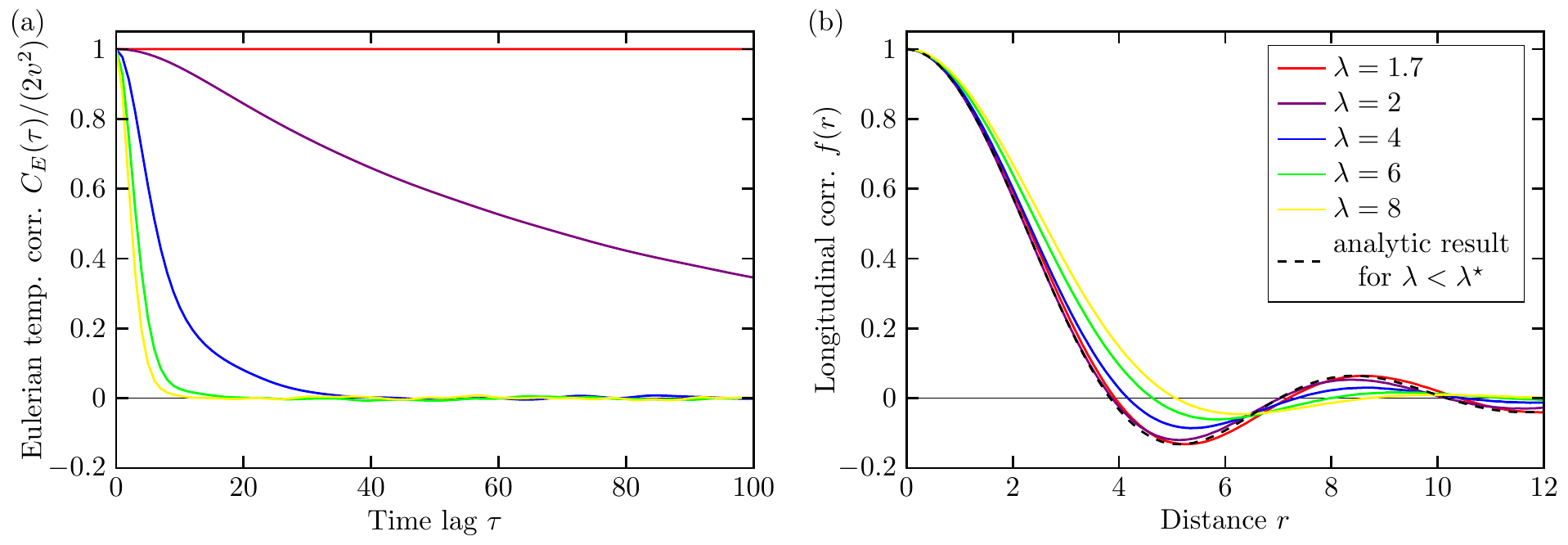}
\caption{\label{fig: Eulerian corr}Eulerian correlations for $a=0.5$ and $b=1.6$. (a) Normalized Eulerian temporal correlation function $C_\mathrm{E}(\tau)$ for different values of $\lambda$. Below the onset of turbulence at $\lambda^\star$, the system is in a stationary state, i.e., temporal correlations do not decay. (b) Longitudinal correlation function $f(r)$ for different values of $\lambda$. Due to the finite-wavelength instability, $f(r)$ shows oscillatory behavior, especially for smaller values of $\lambda$. The longitudinal correlation function below the onset of turbulence at $\lambda^\star$, $f(r)$ can be calculated analytically for a stationary square vortex lattice, see Appendix~\ref{app: vortex lattice}. The	result is included as black dashed line.}
\end{figure}

\begin{figure}
\includegraphics[width=0.99\linewidth]{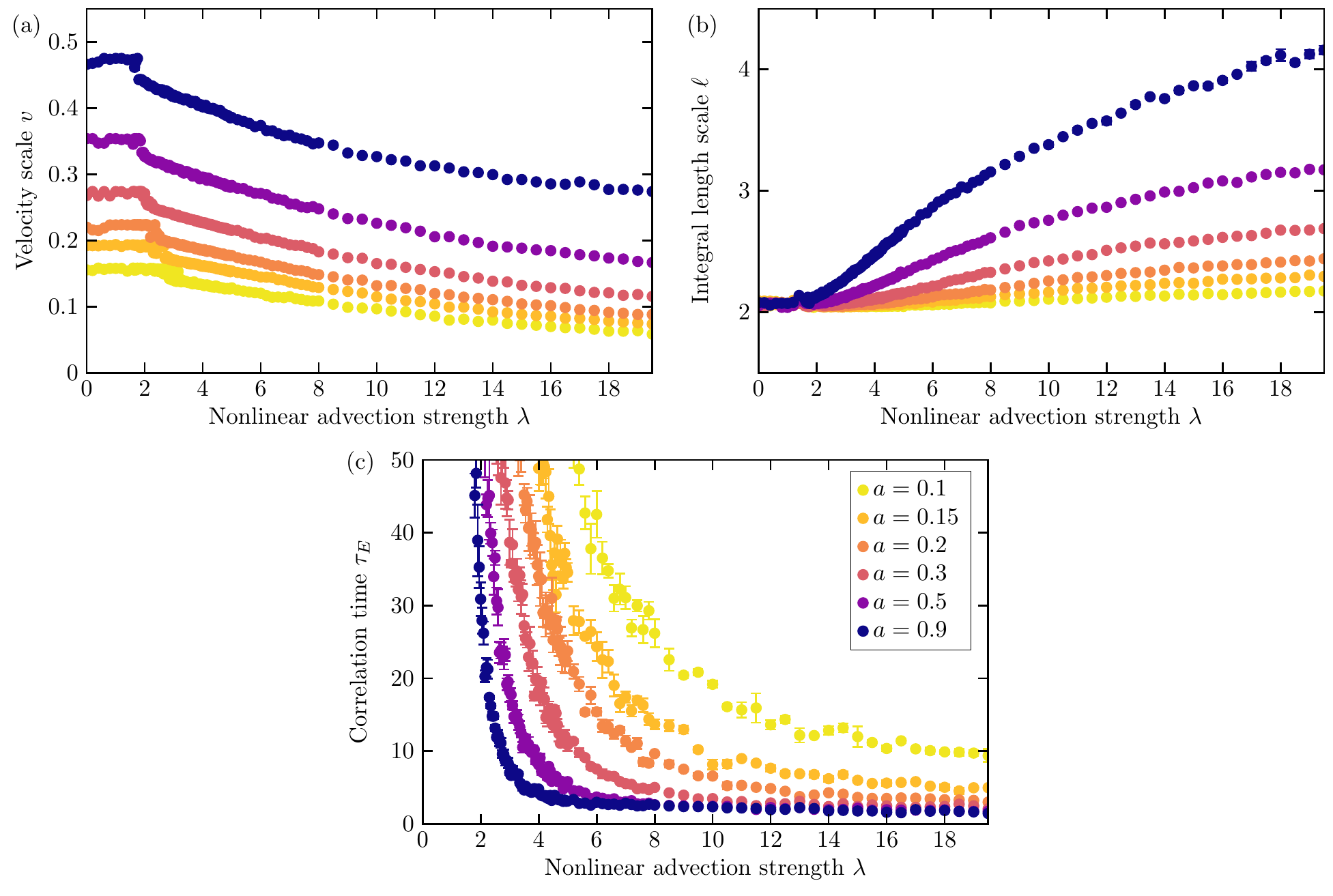}
\caption{\label{fig: Eulerian quan}Quantities calculated in the Eulerian frame of reference for $b=1.6$. (a) Velocity scale $v$ as a function of $\lambda$ for different values of $a$. Above the transition to turbulence, which occurs at $\lambda^\star \approx 1.7 \dots 2.8$ (dependent on $a$), $v$ decreases with increasing $\lambda$. (b) The integral length scale $\ell$ as a function of $\lambda$ for different values of $a$. Below $\lambda^\star$, we observe $\ell = 2$, consistent with a square vortex lattice, see Appendix~\ref{app: vortex lattice}. In the turbulent state, energy is transferred to larger structures, resulting in $\ell$ increasing with $\lambda$. (c) Eulerian correlation time $\tau_\mathrm{E}$ as a function of $\lambda$ for different values of $a$. Below the onset of turbulence at $\lambda^\star$, the system is stationary, which means $\tau_\mathrm{E}$ must diverge. Above $\lambda^\star$, the dynamics gets faster with increasing $\lambda$, resulting in decreasing $\tau_\mathrm{E}$.}
\end{figure}

The Eulerian correlation function is obtained as the trace of the correlation tensor and, in two dimensions, is related to the time-lag-dependent energy spectrum via
\begin{equation}
\label{eq: Eulerian correlation function two-time energy spectrum 2D}
C_\mathrm{E}(r,\tau) = C^\mathrm{E}_{ii}(\mathbf{r},\tau) = 2 \int_0^\infty dk\, \tilde{E}(k,\tau) J_0(kr)\, ,
\end{equation}
where $J_0(x)$ denotes the zeroth-order Bessel function of first kind.
Here, we have again assumed isotropic turbulence.
Correspondingly, $\tilde{E}(k,\tau)$ can be written as
\begin{equation}
\label{eq: two-time energy spectrum 2D Eulerian correlation function}
\tilde{E}(k,\tau) = \frac{1}{2} \int_0^\infty dr\, C_\mathrm{E}(r,\tau) k r J_0(kr)\, .
\end{equation}
The correlation function $C^\mathrm{E}(r,\tau)$ includes both spatial and temporal correlations.
To characterize temporal correlations, we set $r=0$ and obtain the Eulerian temporal correlation function $C_\mathrm{E}(\tau)$,
which is related to the time-lag-dependent energy spectrum $\tilde{E}(k,\tau)$ via
\begin{equation}
\label{eq: temporal Eulerian correlation function from energy spectrum}
C_\mathrm{E}(\tau) = 2 \int_0^\infty dk\, \tilde{E}(k,\tau)\, .
\end{equation}
The spatial correlations, in turn, are characterized by the instantaneous correlation tensor $C^\mathrm{E}_{ij}(\mathbf{r},\tau = 0) = C^\mathrm{E}_{ij}(\mathbf{r})$.
In isotropic, incompressible flows, the full tensor $C^\mathrm{E}_{ij}(\mathbf{r})$ can be written in terms of the longitudinal correlation function $f(r)$ [defined in Eq.~(\ref{eq: longitudinal correlation function})] via~\cite{davidson2015turbulence}
\begin{equation}
\label{eq: relation f and full corr}
C^\mathrm{E}_{ij}(\mathbf{r}) =  v^2 \bigg( \delta_{ij} \frac{\partial}{\partial r} \big[ r f(r)\big] - \frac{r_i r_j}{r} \frac{\partial f(r)}{\partial r} \bigg).
\end{equation}
The scalar spatial Eulerian correlation function $C_\mathrm{E}(r)$ can be calculated via contraction, i.e., $C_\mathrm{E}(r) = C^\mathrm{E}_{ii}(\mathbf{r})$ and is related to $f(r)$ by
\begin{equation}
\label{eq: spatial Eulerian velocity correlation f}
C_\mathrm{E}(r) = \frac{v^2}{r} \frac{\partial}{\partial r} \big[ r^2 f(r)\big]\, .
\end{equation}
Fig.~\ref{fig: Eulerian corr} shows the Eulerian temporal correlation functions $C_\mathrm{E}(\tau)$ and the longitudinal spatial correlation function $f(r)$ at different values of nonlinear advection strength $\lambda$.
Further, in Fig.~\ref{fig: Eulerian quan}, different quantities calculated in the Eulerian framework are plotted as a function of $\lambda$.
Fig.~\ref{fig: Eulerian quan}(a) shows the velocity scale $v$, (b) the integral length scale $\ell$, and (c) the Eulerian correlation time $\tau_\mathrm{E}$, all defined in the main text.

\section{Lagrangian correlations}
\label{app: Lagrangian correlations}

\begin{figure}
\centering
\includegraphics[width=0.85\linewidth]{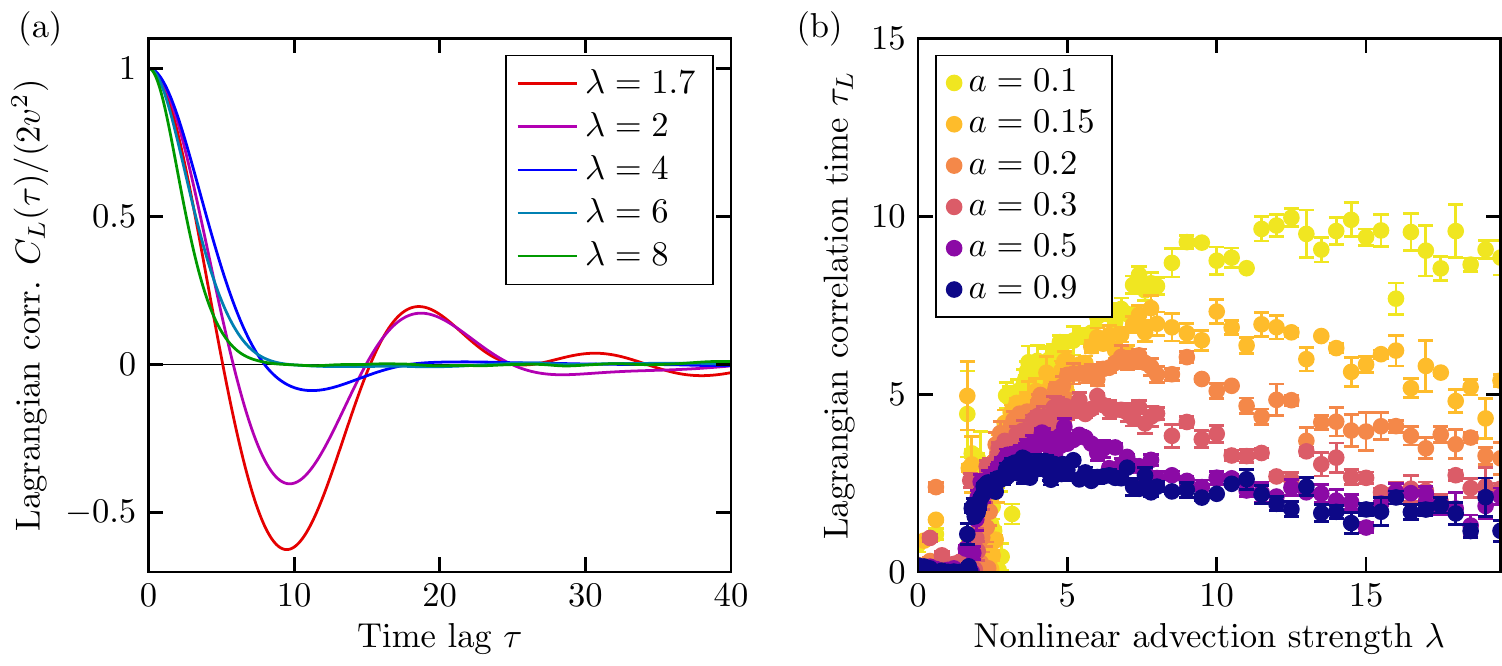}
\caption{\label{fig: Lagrangian corr}(a) Normalized Lagrangian correlation function $C_\mathrm{L}$ as a function of time lag $\tau$ for different values of $\lambda$. The other parameters are $a=0.5$ and $b=1.6$. (b) Lagrangian correlation time $\tau$ as a function of $\lambda$ for different values of $a$ and $b=1.6$.}
\end{figure}

As the tracers sample the Lagrangian statistics of the flow field, we can calculate the Lagrangian correlation function $C_\mathrm{L}(\tau)$ from the trajectories of the tracers via
\begin{equation}
\label{eq: Lagrangian velocity correlation}
C_\mathrm{L}(\tau) = \frac{1}{N} \sum_{i=1}^N \mathbf{u}_i(t_0) \cdot \mathbf{u}_i(t_0+\tau) \, ,
\end{equation}
where $\tau$ is a time lag and $\mathbf{u}_i(t)$ denotes the velocity of tracer $i$ at time $t$.
Analogously to the Eulerian correlation time $\tau_\mathrm{E}$, we calculate the Lagrangian correlation time $\tau_\mathrm{L}$ as
\begin{equation}
\label{eq: Lagrangian correlation time}
\tau_\mathrm{L} = \frac{1}{2 v^2}\int_0^\infty C_\mathrm{L}(t) \ dt \, .
\end{equation}
Fig.~\ref{fig: Lagrangian corr} shows $C_\mathrm{L}(\tau)$ in (a) and $\tau_\mathrm{L}$ as a function of nonlinear advection strength $\lambda$ in (b).
Note that even in the stationary vortex lattice state for nonlinear advection strength $\lambda$ below the thresold value $\lambda^\star$, $C_\mathrm{L}(\tau)$ exhibits oscillatory behavior.
Here, however, we observe $\tau_\mathrm{L} = 0$.
This is because the closed trajectories of the tracers posses different periods depending on their initial position within vortices. 
As there are infinitely many different initial positions, averaging over all tracers results in the integral in Eq.~(\ref{eq: Lagrangian correlation time}) vanishing.
The diffusion coefficient as introduced in the main text is directly related to the Lagrangian correlation time $\tau_\mathrm{L}$ via~\cite{bakunin2008turbulence,davidson2015turbulence}
\begin{equation}
\label{eq: diff coeff Lagrange}
D = \tau_\mathrm{L} v^2\, .
\end{equation}
Increasing the nonlinear advection strength $\lambda$ above the regime where the diffusion constant $D$ is maximal, compare Fig.~\ref{fig: diffusion}(a), we find that the spatial structure of the flow field becomes increasingly unimportant for transport.
Here, the Lagrangian correlation time $\tau_\mathrm{L}$ approaches the Eulerian correlation time $\tau_\mathrm{E}$, until we have $\tau_\mathrm{L} \approx \tau_\mathrm{E}$, which is valid for large $\lambda$ (small Kubo number $K \ll 1$).
In this regime, often denoted as quasilinear~\cite{isichenko1992percolation,vlad2004lagrangian,padberg2007lagrangian}, the diffusion coefficient can be approximated as 
\begin{equation}
\label{eq: diff coeff approx Euler}
D = \tau_\mathrm{E} v^2 \, ,
\end{equation}
or, in nondimensionalized form $D/(\ell v) \approx K$, as is shown in Fig.~\ref{fig: diffusion}(c).

\section{Kubo number as a function of nonlinear advection strength}
\label{app: sweeping}

As observed in Fig.~\ref{fig: diffusion}(b), there is an approximately linear relation between the inverse Kubo number $K^{-1}$ and the nonlinear advection strengh $\lambda$ for values of $\lambda$ sufficiently above the transition to turbulence.
We can understand this linearity based on Kraichnan's random sweeping hypothesis~\cite{kraichnan1964kolmogorov,tennekes1975eulerian}, which was originally used in the context of conventional inertial turbulence. 
Following the reasoning of Kraichnan~\cite{kraichnan1964kolmogorov,wilczek2012wave}, we assume a small-scale velocity field $\mathbf{v}$ that is passively advected by a random sweeping velocity $\mathbf{v}_\mathrm{s}$ with variance $v_\mathrm{s}^2 = \langle\mathbf{v}_\mathrm{s}^2\rangle/2$.
Thus, in Fourier space, we have
\begin{equation}
\label{eq: dynamic equation linearized sweeping Fourier space Appendix}
\partial_t \mathbf{v}(\mathbf{k},t) = - i \lambda \mathbf{k} \cdot \mathbf{v}_\mathrm{s} \mathbf{v}(\mathbf{k},t)\, .
\end{equation}
Note that nonlinear advection in our model is scaled by the coefficient $\lambda$, see Eq.~(\ref{eq: dynamic equation}). Hence, we have also added $\lambda$ in Eq.~(\ref{eq: dynamic equation linearized sweeping Fourier space Appendix}).
We can readily solve Eq.~(\ref{eq: dynamic equation linearized sweeping Fourier space Appendix}), yielding
\begin{equation}
\label{eq: solution sweeping Fourier space}
\mathbf{v}(\mathbf{k},t) = \exp(- i \lambda \mathbf{k} \cdot \mathbf{v}_\mathrm{s} t) \mathbf{v}(\mathbf{k},0)\, .
\end{equation}
This result can be used to connect the time-lag-dependent energy spectrum $\tilde{E}(k,\tau)$ to the instantaneous spectrum $\tilde{E}(k)$, which means we can evaluate the temporal Eulerian correlation function given by Eq.~(\ref{eq: temporal Eulerian correlation function from energy spectrum}).
The detailed calculation is given in~\cite{wilczek2012wave} for inertial Navier-Stokes turbulence.
Taking into account the additional coefficient $\lambda$ is straightforward and we find
\begin{equation}
\label{eq: two-time energy spectrum connected to instantaneous energy spectrum}
\tilde{E}(k,\tau) = \tilde{E}(k) \langle \exp(- i \lambda \mathbf{k} \cdot \mathbf{v}_\mathrm{s}\tau) \rangle,
\end{equation}
where $\langle \dots \rangle$ denotes an ensemble average over different realizations of the sweeping velocity field $\mathbf{v}_\mathrm{s}$.
Here, we assume a Gaussian ensemble distribution with bivariate variables $v_{s,x}$ and $v_{s,y}$ and a covariance tensor that is diagonal.
Under these assumptions, we can further evaluate the time-lag-dependent energy spectrum, which yields 
\begin{equation}
\label{eq: two-time energy spectrum connected to instantaneous energy spectrum 2 wavenumber}
\tilde{E}(k,\tau) = \tilde{E}(k) \exp\Big(-\frac{1}{2}\lambda^2 k^2 v_\mathrm{s}^2 \tau^2\Big)\, .
\end{equation}
Inserting Eq.~(\ref{eq: two-time energy spectrum connected to instantaneous energy spectrum 2 wavenumber}) into Eq.~(\ref{eq: Eulerian correlation function two-time energy spectrum 2D}) yields
\begin{equation}
\label{eq: Eulerian correlation function energy spectrum 2D}
C_\mathrm{E}(r,\tau) = 2 \int_0^\infty dk\, \tilde{E}(k) J_0(kr) \exp\Big(-\frac{1}{2}\lambda^2 k^2 v_\mathrm{s}^2 \tau^2\Big)\, .
\end{equation}
Thus, we can evaluate the Eulerian correlation function $C_\mathrm{E}(r,\tau)$ if we have knowledge of the energy spectrum $\tilde{E}(k)$, which is completely determined by the statistics of the velocity field $\mathbf{v}$.
In particular, the Eulerian temporal correlation function is obtained by setting $r=0$, i.e., 
\begin{equation}
\label{eq: Eulerian temporal correlation sweeping}
C_\mathrm{E}(\tau) =  2 \int_{0}^{\infty} dk \; \tilde{E}(k)\exp\Big(-\frac{1}{2}\lambda^2 k^2 v_\mathrm{s}^2 \tau^2\Big) \, .
\end{equation}
The Eulerian correlation time $\tau_\mathrm{E}$ is calculated via integration, see Eq.~(\ref{eq: integral length scale and Eulerian correlation time}), yielding
\begin{equation}
\label{eq: Eulerian correlation time sweeping 1}
\tau_\mathrm{E} = v^{-2} \int_{0}^{\infty} d\tau \int_{0}^{\infty} dk \; \tilde{E}(k)\exp\Big(-\frac{1}{2}\lambda^2 k^2 v_\mathrm{s}^2 \tau^2\Big) \, .
\end{equation}
The time integral can be evaluated, which yields
\begin{equation}
\label{eq: Eulerian correlation time sweeping 2}
\tau_\mathrm{E} = \frac{\sqrt{2\pi}}{2 \lambda v_\mathrm{s} v^{2}}\int_{0}^{\infty} \frac{\tilde{E}(k)}{k} dk  \, .
\end{equation}
Inserting this into Eq.~(\ref{eq: Kubo number}), we can calculate the Kubo number $K$ via
\begin{equation}
\label{eq: Kubo number sweeping}
K = \frac{\sqrt{2\pi}}{ 2 \lambda v_\mathrm{s} \ell v }\int_{0}^{\infty} \frac{\tilde{E}(k)}{k} dk  \, .
\end{equation}
The above integral is directly connected to the integral length scale~\cite{davidson2015turbulence}.
In two dimensions, we have
\begin{equation}
\label{eq: integral energy spectrum 2D}
\int_{0}^{\infty} \frac{\tilde{E}(k)}{k} dk = \frac{\ell v^2}{2}  \, ,
\end{equation}
which is shown in Appendix~\ref{app: connection energy spectrum integral length}.
Thus, we find the following simple equation for the inverse Kubo number $K^{-1}$,
\begin{equation}
\label{eq: inverse Kubo number sweeping 2D}
K^{-1} =  \frac{4}{\sqrt{2\pi}}\frac{\lambda v_\mathrm{s}}{v} \, .
\end{equation}
As the fluctuations of the sweeping velocity field $\mathbf{v}_\mathrm{s}$ ultimately stem from the field $\mathbf{v}$, it is reasonable to assume that the component-wise variance $v_\mathrm{s}^2$ of the sweeping velocity field is proportional to the component-wise variance of the small-scale velocity field given by $\langle v_x^2\rangle = \langle v_y^2\rangle = v^2$, i.e.,
\begin{equation}
\label{eq: sweeping velocity}
v_\mathrm{s} = c \, v\, ,
\end{equation}
where $c$ is a coefficient to be determined via comparison with the numerical results.
Thus, we find the simple scaling
\begin{equation}
\label{eq: inverse Kubo number sweeping 2D 2}
K^{-1} = \frac{4c\lambda}{\sqrt{2\pi}} \propto \lambda \, .
\end{equation}
This linear dependence can be observed in Fig.~\ref{fig: diffusion}(b) for larger values of $\lambda$.

\begin{figure}
\includegraphics[width=0.48\linewidth]{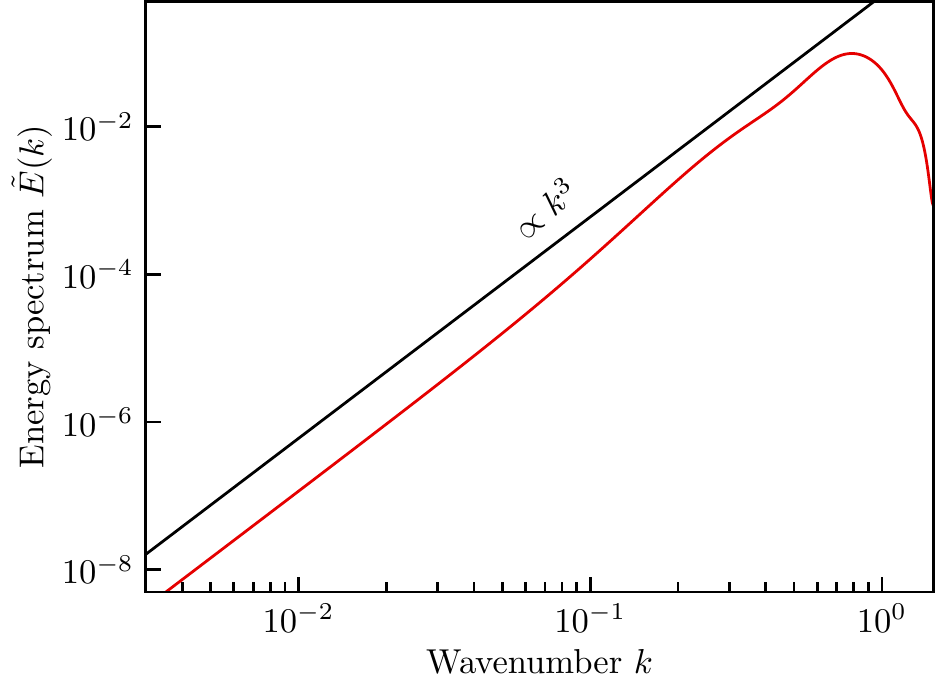}
\caption{\label{fig: energy spectrum}Energy spectrum $\tilde{E}(k)$ as a function of the wavenumber for $\lambda=10$, $a=0.5$ and $b=1.6$. The spectrum increases as $\propto k^3$, shows a clear maximum, and then decreases exponentially for larger $k$.}
\end{figure}

\begin{figure}
\includegraphics[width=0.99\linewidth]{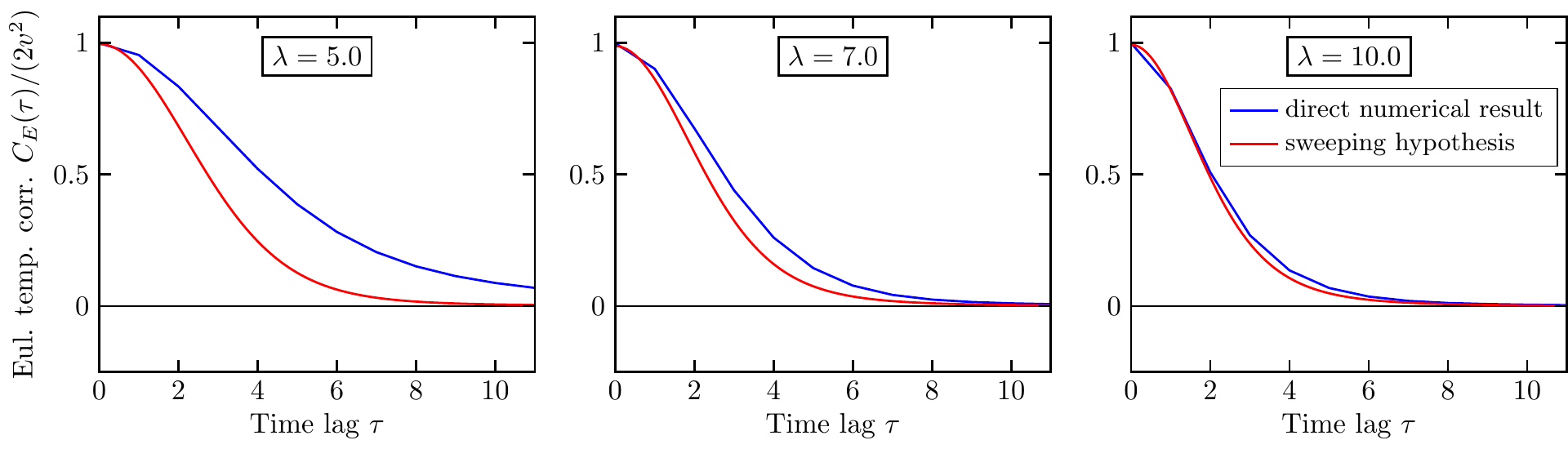}
\caption{\label{fig: correlation function sweeping}Correlation function $C_\mathrm{E}(\tau)$ for $\lambda=5$, $\lambda=7$ and $\lambda=10$. The blue curve shows the correlation function calculated directly from the numerical results. The red curve shows the correlation function calculated via Eq.~(\ref{eq: Eulerian temporal correlation sweeping}) with $c=0.33$. The energy spectrum $\tilde{E}(k)$ was calculated from the spatial correlation function $f(r)$ determined from the numerical simulations. The other parameters are $a=0.5$ and $b=1.6$.}
\end{figure}

To further test the sweeping hypothesis, we can directly compare the temporal correlation function $C_\mathrm{E}(\tau)$ calculated numerically by looking at time-resolved data [using Eq.~(\ref{eq: Eulerian temporal correlation function})] with $C_\mathrm{E}(\tau)$ calculated via Eq.~(\ref{eq: Eulerian temporal correlation sweeping}).
To this end, we have to determine the energy spectrum $\tilde{E}(k)$, e.g., via calculation from the longitudinal correlation function $f(r)$ (also determined numerically) using Eqs.~(\ref{eq: two-time energy spectrum 2D Eulerian correlation function}) for $\tau=0$ and (\ref{eq: spatial Eulerian velocity correlation f}).
Fig.~\ref{fig: energy spectrum} shows $\tilde{E}(k)$ for $\lambda=10$, $a=0.5$ and $1.6$.
As in~\cite{bratanov2015new}, we observe that $\tilde{E}(k)$ increases as $\propto k^3$ for small $k$, shows a clear maximum and then decreases exponentially for larger $k$.
Inserting $\tilde{E}(k)$ into Eq.~(\ref{eq: Eulerian temporal correlation sweeping}) and integrating, we can calculate $C_\mathrm{E}(\tau)$ on the basis of the sweeping hypothesis and test whether it gives an accurate estimate.
Fig.~\ref{fig: correlation function sweeping} shows the comparison with direct numerical results for $C_\mathrm{E}(\tau)$ at different values of $\lambda$.
As in Fig.~\ref{fig: diffusion}(b), we have set $c=0.33$, which gives quite accurate results for larger values of $\lambda$, e.g., at $\lambda = 10$.
This is consistent with the linear behavior observed in Fig.~\ref{fig: diffusion}(b) for large advection strength.

\section{Connection between energy spectrum and integral length scale}
\label{app: connection energy spectrum integral length}

The integral in Eq.~(\ref{eq: Kubo number sweeping}) is directly connected to the integral length scale.
To show this, we first insert Eq.~(\ref{eq: two-time energy spectrum 2D Eulerian correlation function}) for $\tau=0$ into the integral, i.e.,
\begin{equation}
\label{eq: integral energy spectrum 2D step 1}
\int_{0}^{\infty}  \frac{\tilde{E}(k)}{k} \,dk  = \frac{1}{2} \int_{0}^{\infty} dk\, \int_0^\infty dr \, C_\mathrm{E}(r) r J_0(kr)  \, .
\end{equation}
As only $J_0(kr)$ depends on $k$, we can integrate over $k$ yielding
\begin{equation}
\label{eq: integral energy spectrum 2D step 2}
\int_{0}^{\infty}  \, \frac{\tilde{E}(k)}{k} \, dk = \frac{1}{2} \int_0^\infty  C_\mathrm{E}(r) \, dr \, .
\end{equation}
The spatial correlation function $C_\mathrm{E}(r)$ can be written in terms of the longitudinal correlation function, see Eq.~(\ref{eq: spatial Eulerian velocity correlation f}), which yields
\begin{equation}
\label{eq: integral energy spectrum 2D step 3}
\begin{aligned}
\int_{0}^{\infty} \, \frac{\tilde{E}(k)}{k} \, dk  = \frac{v^2}{2} \int_0^\infty \left( 2 f(r) + r \frac{\partial f(r)}{\partial r} \right) dr \\  
= v^2 \int_0^\infty f(r) \, dr +  \frac{v^2}{2} \int_0^\infty r \frac{\partial f(r)}{\partial r} \, dr \, .
\end{aligned}
\end{equation}
The second integral on the right-hand side can be computed via integration by parts, i.e.,
\begin{equation}
\label{eq: integral energy spectrum 2D step 3 intermediate}
\int_0^\infty r \frac{\partial f(r)}{\partial r} \, dr =  \Big[ r f(r) \Big]_0^\infty - \int_0^\infty f(r) \, dr \, .
\end{equation}
The boundary term vanishes because $f(r)$ decays faster than linear for large $r$.
Thus, inserting Eq.~(\ref{eq: integral energy spectrum 2D step 3 intermediate}) into Eq.~(\ref{eq: integral energy spectrum 2D step 3}) yields 
\begin{equation}
\label{eq: integral energy spectrum 2D step 4}
\int_{0}^{\infty} \frac{\tilde{E}(k)}{k} \, dk  = \frac{v^2}{2} \int_0^\infty f(r) \, dr \, .
\end{equation}
With the definition of the integral length scale $\ell$, see Eq.~(\ref{eq: integral length scale and Eulerian correlation time}), we finally have
\begin{equation}
\label{eq: integral energy spectrum 2D step 5}
\int_0^\infty  \frac{\tilde{E}(k)}{k} \, dk = \frac{\ell v^2}{2}\, .
\end{equation}

\section{Generality of the results}
\label{app: variation of b}

\begin{figure}
\includegraphics[width=0.99\linewidth]{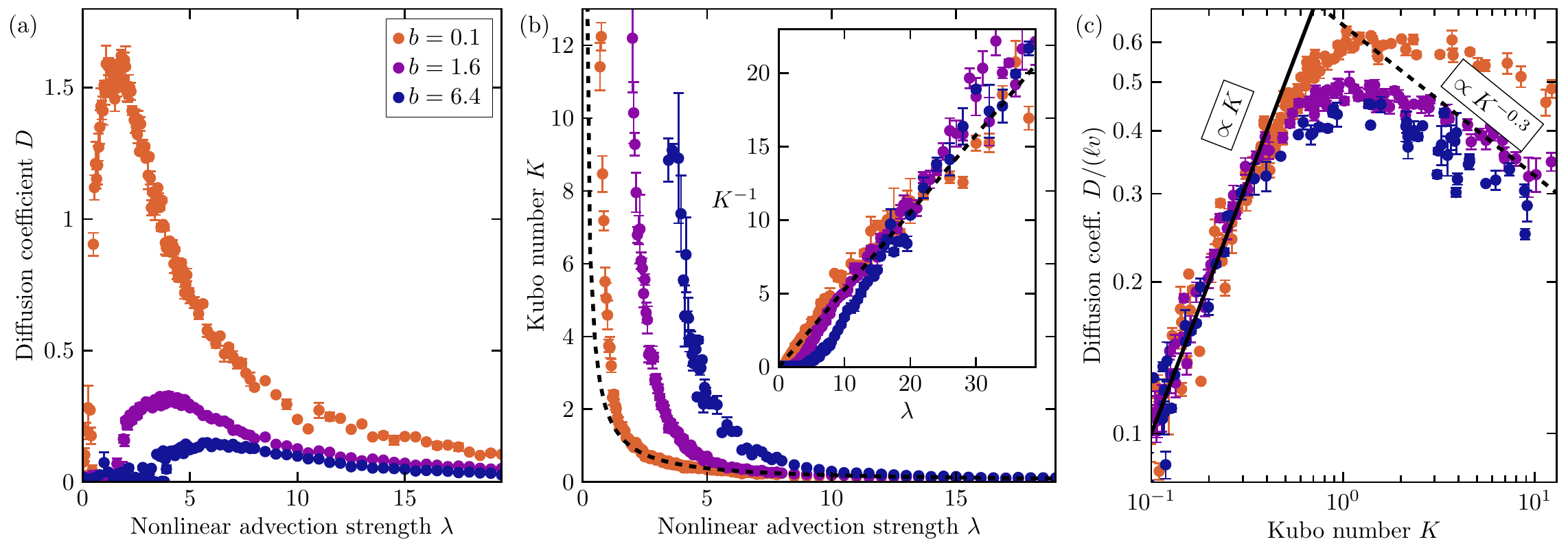}
\caption{\label{fig: varying b}(a) Diffusion coefficient $D$ as a function of nonlinear advection strength $\lambda$ for different values of $b$. (b) Kubo number $K$ as a function of $\lambda$ for different values of $b$. $K$ is finite in the mesoscale-turbulent state for $\lambda > \lambda^\star$ and diverges when approaching the stationary vortex lattice state for $\lambda < \lambda^\star$. Inset: inverse Kubo number $K^{-1}$ as a function of $\lambda$, which shows linear behavior for larger values of $\lambda$. The dashed line gives the result obtained via Kraichnan's random sweeping hypothesis with $c = 0.33$. (c) Dimensionless diffusion coefficient $D/(\ell v)$ as a function of Kubo number $K$. The scaling in the different regimes is given as solid and dashed line, respectively. The remaining parameter is $a=0.5$.}
\end{figure}

In the main text of this article, we varied the value of the coefficient of the linear term, $a$, and left the coefficient of the cubic term unchanged, $b = 1.6$.
In order to test the generality of our results, we performed additional numerical calculations varying $b$, while keeping $a$ constant.
The results are shown in Fig.~\ref{fig: varying b} for three values $b = 0.1$, $b = 1.6$, and $b = 6.4$ (and $a = 0.5$).
Note that the results for $a = 0.5$ and $b = 1.6$ are also presented in the main text.

Investigating diffusion coefficient and Kubo number, we find the same qualitative behavior for all values of $b$.
After an initial trapping regime in the vortex lattice for small advection strength $\lambda$, the diffusion coefficient $D$ increases rapidly above the transition occurring at $\lambda^\star$ until a maximum is reached.
After this regime of optimal transport, the diffusion coefficient decreases.
We further note that, for all values of $\lambda$, the diffusion coefficient is larger, the smaller $b$.
This is because the cubic term is responsible for the saturation of the emerging patterns. Thus, decreasing $b$ increases the velocity scale $v$, which directly increases the diffusion coefficient, see Eq.~(\ref{eq: diff coeff Lagrange}).
Regarding the Kubo number $K$, we find a linear relation between $K$ and advection strength $\lambda$, not only at $b = 1.6$ (main text), but also for the other values of $b$.
Remarkably, the coefficient $b$ has no impact on the proportionality constant $c = 0.33$, see Fig.~\ref{fig: varying b}(b).
Further, Fig.~\ref{fig: varying b}(c) shows the scaling behavior of the non-dimensionalized diffusion coefficient $D/(\ell v)$ as a function of $K$.
As in Fig.~\ref{fig: diffusion}, we observe a regime $D \propto K$ for small $K$, where transport is completely determined by the temporal correlations.
For large $K$, the data indicate again a regime where $D \propto K^{-0.3}$, which is a result of the frequent trapping of tracers in persistent vortices.
However, note that the data points for the dimensionless diffusion coefficient $D/(\ell v)$ at different values of $b$ do not collapse onto one curve for $K \gg 1$.
In contrast, for constant $b$ and varying $a$, this is the case, compare Fig.~\ref{fig: diffusion}(c).
This observation suggests that the cubic nonlinearity, which is scaled by $b$ (see Eq.~(1) in the main text), has a strong quantitative impact on transport in particular in the slowly evolving system ($K \gg 1$) close to the transition between turbulence and vortex lattice for smaller values of $\lambda$ close to $\lambda^\star$. 
However, our results indicate that the coefficient $b$ does not impact the scaling exponents.
To conclude, the transport properties of the microswimmer suspension described in detail in the main text and summarized above are quite general for large regions of parameter space.

%

\end{document}